\documentstyle[12pt]{article}
\newcommand{\newsection}{    % Numeration of eqs. is automatic
\setcounter{equation}{0}
\section}
\def\appendix#1{
\addtocounter{section}{1}
\setcounter{equation}{0}
\renewcommand{\thesection}{\Alph{section}}
\section*{Appendix \thesection\protect\indent #1}
\addcontentsline{toc}{section}{Appendix \thesection\ \ \ #1}
}
\newcommand{\rf}[1]{(\ref{#1})}

\def\be{\begin{equation}}
\def\ee{\end{equation}}
\newcommand{\beq}{\begin{equation}}
\newcommand{\eeq}{\end{equation}}
\newcommand{\bea}{\begin{eqnarray}}
\newcommand{\eea}{\end{eqnarray}}

\renewcommand{\l}{\lambda}

\newcommand{\th}{\theta}
\newcommand{\tht}{\tilde{\theta}}

\newcommand{\non}{\nonumber}

\newcommand{\tr}{{\,\rm tr}\:}
\newcommand{\La}{\Lambda}
\newcommand{\p}{{\cal P}}

\newcommand{\hs}{\hspace{0.7cm}}
\begin{document}
\topmargin 0pt
\oddsidemargin 5mm
\headheight 0pt
\headsep 0pt
\topskip 9mm

\hfill SPhT/94-102
\addtolength{\baselineskip}{0.20\baselineskip}
\begin{center}
\vspace{26pt}
{\large \bf Multi-loop correlators for rational theories of 2D gravity
from the generalized Kontsevich models}
\vspace{26pt}
\newline
C.\ Kristjansen\footnote{On leave of absence from NORDITA, Blegdamsvej 17,
DK-2100 Copenhagen \O, Denmark} \\
\vspace{6pt}
Service de Physique Th\'{e}orique de Saclay \\
F-91191 Gif-sur-Yvette Cedex, France \\
\end{center}
\vspace{20pt}
\begin{center}
{\bf Abstract}
\end{center}
We introduce a parametrization of the coupling constant space of the
generalized Kontsevich models in terms of a set of moments equivalent to those
introduced
recently in the context of topological gravity. For the simplest generalization
of the Kontsevich model we express the moments as elementary functions of the
susceptibilities and the eigenvalues of the external field. We furthermore use
the moment technique to derive a closed expression for the genus zero
multi-loop correlators for $(3,3m-1)$ and $(3,3m-2)$ rational matter fields
coupled to
gravity. We comment on the relation between the two-matrix model and the
generalized Kontsevich models.
\newpage

\newsection{Introduction}
Several types of matrix models have proven to have a singularity structure to
which a scaling behaviour characteristic of two-dimensional conformal field
theories coupled to gravity can be associated. The simplest example is the
generic 1-matrix model which possesses singular points which, when approached
by the double scaling procedure~\cite{dsl90}, give rise to the scaling
behaviour characteristic of $(2,2m-1)$ conformal matter coupled to
gravity~\cite{conf90}.
However, the generic 1-matrix model is not the most economic way of studying
the series of $(2,2m-1)$ theories because in the vicinity of any of its $m$'th
multi-critical points there will be subleading singularities present. The most
economic way  of studying the interaction of $(2,2m-1)$ matter with gravity is
by using the Kontsevich model~\cite{Kon91,Wit91}. In the parameter space of
this model one recovers all the $m$'th multi-critical regimes of the 1-matrix
model but without the presence of any subleading singularities.

Recently it has been shown that the two-matrix model possesses critical points
capable of describing the interaction of any $(p,q)$ rational matter field with
2-dimensional quantum gravity~\cite{DKK93}. As in the previous case, due to the
presence of subleading singularities, the generic 2-matrix model is not the
optimal tool for studying these interactions. The optimal line of action would
be to consider a model which possessed singular points of the same type
as those of the two-matrix model but were deprived of any subleading
singularities. Although the situation is not completely as clear as in the
1-matrix case, the generalized Kontsevich models seem to be the models we
are looking for~\cite{Kon91,AM92,IZ92,KMMMZ92}. The partition functions of
these models read in the normalization of reference~\cite{IZ92}
\beq
Z_p^N(\La)=\frac{\int d M\exp \left( \frac{i^{p^2+1}}{2(p+1)}\tr\left[
\left(M+(-i)^{p+1}\La \right)^{p+1}\right]_{>lin}\right)}
{\int d M \exp\left(-\frac{1}{4}\tr \left[ \sum_{k=0}^{p-1} M \La^k M
\La^{p-1-k}
\right] \right)}
\label{genpart}
\eeq
where the integration is over $N\times N$ hermitian matrices and where
$\Lambda$
is an external field. The subscript
$>lin$ means that only terms of degree larger than or equal to two in
$M$ should be taken into account. The usual Kontsevich model
is recovered for $p=2$. All matrix integrals of the type~\rf{genpart} have an
expansion in powers of the
traces, $\tr \La^{-n}$, but $Z_p^N(\La)$ is independent of
$\tr \La^{-np}$~\cite{Kon91,IZ92}.
Expressed in terms of the remaining traces $Z_p^N(\La)$ is known to fulfill a
set of $W_p$
constraints or equivalently to be a $\tau$-function of the $p^{th}$ reduction
of the KP hierarchy, the KdV$_p$ hierarchy, and fulfill the string equation
$L_{-1}Z_p^N(\La)=0$~\cite{IZ92,KMMMZ92}. Hence $Z_p^N(\La)$ should be capable
of describing the coupling of rational matter fields of the type
$(p,pm-1),\ldots(p,pm-(p-1))$ to two-dimensional quantum gravity.

In the present paper we will study the $p=3$ version of the model~\rf{genpart}
with the aim of extracting information  about the interaction of $(3,3m-1)$
and $(3,3m-2)$ matter fields with gravity. In the case of the ordinary
Kontsevich model as well as in the case of the generic 1-matrix model it
proved convenient to parametrize the coupling constant space by a set of
so-called moment variables~\cite{IZ92,ACKM93}. After having fixed the notation
in section~\ref{model} we will in section~\ref{solution} introduce the
appropriate moment variables for the $p=3$ version of~\rf{genpart} and scetch
how the idea can be generalized to the generic case. Our moment variables are
exactly equivalent to those introduced in reference~\cite{EYY94} in the context
of topological gravity. However, in the matrix model framework one can express
the moments explicitly in terms of elementary functions of the susceptibilities
and the eigenvalues of the external matrix. We hence obtain an expression for
the genus zero contribution to the free energy of the $p=3$ model where the
singularities are clearly exposed. In section~\ref{macloop} we use the moment
technique to study the macroscopic loops of $(3,3m-1)$ and $(3,3m-2)$ matter
fields coupled to quantum gravity. We derive a closed expression for the
$n$-loop correlator at genus zero thereby generalizing the expression obtained
earlier for the $(2,2m-1)$ case in references~\cite{AJM90,MSS91}. Our results
allow us to extract several characteristics of the multi-loop correlators of
the generic $(p,q)$-model.

In section~\ref{conclusion} we comment on the exact relation between the
two-matrix model and the generalized Kontsevich models as well as the relation
of the present formalism to that of strings with discrete target
space~\cite{Kos91}.

\newsection{The model \label{model} }
The model that we will consider is the $p=3$ version of the partition function
given in~\rf{genpart}
\beq
Z_N(\La)=e^{F_N(\La)}=\int_{N\times N} d\mu (M)
\exp \left\{-\frac{1}{2} \tr \left( M^3 \La+\frac{1}{4} M^4\right) \right\}
\label{partition}
\eeq
where the measure reads
\beq
\mu(M)=\frac{dM \exp\left(-\frac{1}{2}\tr \left[\La^2 M^2+
\frac{1}{2}\La M\La M \right] \right)}
{\int dM \exp \left(-\frac{1}{2}\tr \left[ \La^2 M^2+
\frac{1}{2}\La M\La M\right] \right)}.
\eeq
We introduce time variables $\{\th_k,\tht_k\}$ for the model by
\beq
\th_k=\frac{1}{3k+1}\tr \La^{-3k-1}, \hs
\tht_k=\frac{1}{3k+2}\tr \La^{-3k-2}, \hspace{0.3cm} k\geq 0. \label{times}
\eeq
Then the fact that $Z_N(\La)$ is a $\tau$-function of the kdV$_3$ hierarchy
can be expressed as
\beq
\frac{\partial Q}{\partial \th_s}=\left[ Q_+^{(3s+1)/3},Q\right], \hs
\frac{\partial Q}{\partial \tht_s}=\left[ Q_+^{(3s+2)/3},Q\right],
\label{pseudo}
\eeq
where
\beq
Q=\left(\frac{\partial}{\partial \th_0}\right)^3+
\frac{3}{2}\left\{u_1,\frac{\partial}{\partial \th_0}\right\}+3u_2
\eeq
\beq
u_1=\frac{\partial^2 F }{\partial \th_0^2}, \hs
u_2=\frac{1}{2}\frac{\partial^2 F}{\partial \th_0 \partial \tht_0}
\eeq
and the constraint $L_{-1} Z_N(\Lambda)$=0
implies that
\bea
\frac{\partial u_1}{\partial \th_0}&=&\frac{2}{3}
\sum_{s\geq 0} (3s+4)\th_{s+1}\frac{\partial u_1}{\partial \th_s}
+\frac{2}{3}\sum_{s\geq 0} (3s+5) \tht_{s+1}\frac{\partial u_1}{\partial
\tht_s}
\label{flow1}\\
\frac{\partial u_2}{\partial \th_0}&=&\frac{2}{3}
\sum_{s\geq 0} (3s+4)\th_{s+1}\frac{\partial u_2}{\partial \th_s}
+\frac{2}{3}\sum_{s\geq 0} (3s+5) \tht_{s+1}\frac{\partial u_2}{\partial
\tht_s}
+\frac{2}{3}.
\label{flow2}
\eea
This information allows us to solve the model to any order in $1/N^2$.
The time variables $\{\th_k,\tht_k\}$ are related to the $t_{n,m},\;n=0,1\;
;m\geq 0$ used by Witten~\cite{Wit91} in the context of topological gravity by
\bea
\th_k=t_{0,k}\; \frac{\rho^{3k+1}}{(3k+1)!!!}\left(\frac{\sqrt{3}}{i}\right)^k,
\hs
\tht_k=t_{1,k}\; \frac{\rho^{3k+2}}{(3k+2)!!!}\left(\frac{\sqrt{3}}{i}\right)
^{k+1}, &&\\
\rho^4=\frac{3^{3/2}}{2i},\hs (3k+m)!!!=(3k+m)(3(k-1)+m)\ldots m,\;\;\; m=1,2.
\non
&&
\eea

\newsection{The solution \label{solution} }
Let us introduce the following notation for the negative part of the pseudo
differential operators entering the equations~\rf{pseudo}
\bea
Q_-^{(3s+1)/3} &\equiv& P_s \partial^{-1}+(Q_s-\frac{1}{2} P_s')\partial^{-2}
+{\cal O}(\partial^{-3}), \label{Q13}\\
Q_-^{(3s+2)/3} & \equiv & \tilde{P}_s \partial^{-1}+
(\tilde{Q}_s-\frac{1}{2}\tilde{P}_s')\partial^{-2}+ {\cal O}(\partial^{-3})
\label{Q23}
\eea
where $\partial=\frac{\partial}{\partial \th_0}$ and primes refer to
differentiation with respect to $\th_0$. Then the flow equations for $u_1$
and $u_2$ can be written as
\bea
\frac{\partial u_1}{\partial \th_s}=P_s',&\hs & \frac{\partial u_1}
{\partial \tht_s}=\tilde{P}_s'
\label{flowu1} \\
\frac{\partial u_2}{\partial \th_s}=Q_s',&\hs & \frac{\partial u_2}
{\partial \tht_s}=\tilde{Q}_s'.
\label{flowu2}
\eea
The functions $\{P_i,\tilde{P}_i,Q_i,\tilde{Q}_i\}$ are polynomials in $u_1$
and $u_2$ and the derivatives of these and determined by
\bea
P_0=u_1,&\hs & Q_0=u_2, \\
\tilde{P}_0=2u_2,&\hs & \tilde{Q}_0=-\frac{1}{6}u_1''-u_1^2
\eea
plus the following set of recursion relations which can be derived in the
standard way
\bea
P_{s+1}'&=&\frac{2}{3}Q_s^{(3)}+2u_1Q_s'+u_1'Q_s+2u_2'P_s+3P_s'u_2, \\
Q_{s+1}'&=&-\frac{1}{18}P_s^{(5)}-\frac{5}{6}u_1 P_s^{(3)}-
\frac{1}{6}u_1^{(3)}P_s
-\frac{5}{4}u_1'P_s''-\frac{3}{4}u_1''P_s' \non\\
&& \mbox{}+3u_2Q_s'+u_2'Q_s-2u_1^2P_s'-2u_1u_1'P_s.
\eea
Let us consider the planar limit $N\rightarrow \infty$. In this limit
we can neglect all higher derivatives in the recursion relations. It is
possible
to show that under these circumstances the polynomials take the following
general form
\bea
P_k&=&(3k+1)!!!\sum_{j=0}^{[k/2]}
\frac{(3j-1)!!!(-1)^j}{(k-2j)!(3j+1)!}\;u_1^{3j+1}u_2^{k-2j},
\label{Pk} \\
Q_k&=&(3k+1)!!!\sum_{j=0}^{[(k+1)/2]}
\frac{(3j-1)!!!(-1)^j}{(k+1-2j)!(3j)!}\;u_1^{3j}u_2^{k+1-2j},
\label{Qk} \\
\tilde{P}_k&=&(3k+2)!!!\sum_{j=0}^{[(k+1)/2]}
\frac{(3j-2)!!!(-1)^j}{(k+1-2j)!(3j)!}\:u_1^{3j}u_2^{k+1-2j},
\label{Ptk}\\
\tilde{Q}_k&=&(3k+2)!!!\sum_{j=0}^{[k/2]}
\frac{(3j+1)!!!(-1)^{j+1}}{(k-2j)!(3j+2)!}\:u_1^{3j+2}u_2^{k-2j}
\label{Qtk}
\eea
where $[a]$ denotes the integer part of $a$.
These polynomials can be shown to fulfill the following relations
\bea
\frac{\partial P_k}{\partial u_1}=(3k+1)Q_{k-1}, &\hs &
\frac{\partial Q_k}{\partial u_1}=-(3k+1)u_1 P_{k-1},
\label{derivative1}\\
\frac{\partial {P}_k}{\partial u_2}=(3k+1)P_{k-1}, &\hs &
\frac{\partial Q_k}{\partial u_2}=(3k+1) Q_{k-1}.
\label{derivative2}
\eea
Similar relations where $(3k+1)$ is replaced by $(3k+2)$ hold for the
polynomials
$\{\tilde{P}_k,\tilde{Q}_k\}$.
Inspired by the form of the flow equations~\rf{flow1} and~\rf{flow2}
let us introduce two sets of moment variables by
\bea
M_k&=&\frac{2}{3}\left\{\sum_{s\geq -1}\th_{s+k}P_s
\frac{(3(s+k)+1)!!!}{(3s+1)!!!}
+\sum_{s\geq -1}\tht_{s+k}\tilde{P}_s \frac{(3(s+k)+2)!!!}{(3s+2)!!!}
\right\},
\label{momentM}\\
J_k&=&\frac{2}{3}\left\{\sum_{s\geq -1}\th_{s+k}Q_s
\frac{(3(s+k)+1)!!!}{(3s+1)!!!}
+\sum_{s\geq -1}\tht_{s+k}\tilde{Q}_s \frac{(3(s+k)+2)!!!}{(3s+2)!!!}
\right\}
\label{momentJ}
\eea
where the polynomials with negative indices are defined by the
relations~\rf{derivative1} and~\rf{derivative2}.
Let us notice that the
relations~\rf{derivative1} and~\rf{derivative2} for the polynomials
$\{P_k,\tilde{P}_k,Q_k,\tilde{Q}_k\}$ imply the following relations between
the moments
\bea
\frac{\partial M_k}{\partial u_1}=J_{k+1},&\hs &
\frac{\partial J_k}{\partial u_1}=(-u_1)M_{k+1},
\label{momentderivative1}\\
\frac{\partial M_k}{\partial u_2}=M_{k+1},&\hs &
\frac{\partial J_k}{\partial u_2}=J_{k+1}.
\label{momentderivative2}
\eea
Now we can write the
flow equations~\rf{flow1} and~\rf{flow2} as
\beq
\frac{\partial}{\partial\th_0}(u_1-M_1)=0,\hs
\frac{\partial}{\partial\th_0}(u_2-J_1)=0.
\eeq
Furthermore one can show that one has in addition
\bea
\frac{\partial}{\partial\th_k}(u_1-M_1)=0,&\hs &
\frac{\partial}{\partial\th_k}(u_2-J_1)=0, \hspace{0.5cm} k\geq 1;\\
\frac{\partial}{\partial\tht_k}(u_1-M_1)=0,&\hs &
\frac{\partial}{\partial\tht_k}(u_2-J_1)=0, \hspace{0.5cm}k\geq 0.
\eea
This can be seen from rewritings of the following type
\beq
\frac{\partial u_1}{\partial \th_k}=\frac{\partial}{\partial \th_0} P_k=
\frac{\partial P_k}{\partial u_1} \frac{\partial u_1}{\partial \th_0}
+\frac{\partial P_k}{\partial u_2}\frac{\partial u_2}{\partial \th_0}
\eeq
followed by application of the relations~\rf{derivative1} and~\rf{derivative2}
as well as the constraints~\rf{flow1} and~\rf{flow2}. Hence $(u_1-M_1)$
and $(u_2-J_1)$ must be constants and since they should vanish for
$\th_i=0,\;\tht_i=0$ one concludes that
\beq
u_1=M_1,\hs u_2=J_1. \label{boundary}
\eeq
These equations give us an implicit expression for $F_0$.
The moment variables~\rf{momentM} and~\rf{momentJ} are identical to those
introduced in reference~\cite{EYY94} for the topological minimal model
associated with the Lie Algebra $A_2$. As in that case the description could
easily be generalised to the case of $A_n$ it is obvious that the strategy
applied here for $p=3$ version of~\rf{genpart} can
easily be generalized to the $p=n$ version. For the $p=n$ version we will have
$(n-1)$ series of time variables, $(n-1)$ susceptibilities and $(n-1)$
relations like~\rf{flow1} and~\rf{flow2}.
The flow equations will be expressed in terms of $(n-1)$ series of pseudo
differential operators which have expansions like~\rf{Q13} and~\rf{Q23} where
now the $(n-1)$ first terms are of importance. Hence we are
led to introduce $(n-1)$ series of polynomials and $(n-1)$ series of moments
each moment being a sum of $(n-1)$ terms in close analogy with~\rf{momentM}
and~\rf{momentJ}. In reference~\cite{EYY94} it was shown that all higher genera
contributions to the free energy can be expressed entirely in terms of the
moments and that for any given model, $A_n$, and given genus, $g$, only a
finite
number of moments appear. This result of course also appears in the
matrix model framework. However, we will not enter into a discussion of this
point. Let us just mention that all higher genera contributions to the free
energy in the case $p=3$ can be found by solving iteratively the genus expanded
version of the flow equation
\beq
\frac{\partial u_1}{\partial \th_1}=\frac{\partial }{\partial \th_0} P_1
=\frac{\partial}{\partial \th_0}\left(4 u_1 u_2 +\frac{2}{3}\frac{\partial^2
u_2}{\partial \th_0^2}\right).
\eeq

In the matrix model framework it is possible to express the moment variables
in terms of elementary functions of the susceptibilities and the eigenvalues
of the external field. For the model~\rf{partition} we find using the explicit
expressions~\rf{Pk}, \rf{Qk}, \rf{Ptk} and~\rf{Qtk} for the polynomials
$P_k$, $Q_k$, $\tilde{P}_k$, $\tilde{Q}_k$ and the definition~\rf{times} of the
time variables~\cite{Han75}
\bea
M_0&=&\frac{2^{4/3}}{3} u_1
\sum_k
\left\{(\l_k^3-3u_2)+\left[(\l_k^3-3u_2)^2+4u_1^3\right]^{1/2}\right\}^{-1/3}
\non\\
&&-
\frac{2^{2/3}}{3}
\sum_k
\left\{(\l_k^3-3u_2)+\left[(\l_k^3-3u_2)^2+4u_1^3\right]^{1/2}\right\}^{1/3}
\label{M0}\\
J_0&=&
\frac{2^{2/3}}{3}(-u_1^2)
\sum_k
\left\{(\l_k^3-3u_2)+\left[(\l_k^3-3u_2)^2+4u_1^3\right]^{1/2}\right\}^{-2/3}
\non \\
&&-
\frac{2^{-2/3}}{3}
\sum_k
\left\{(\l_k^3-3u_2)+\left[(\l_k^3-3u_2)^2+4u_1^3\right]^{1/2}\right\}^{2/3}
\label{J0}
\eea
\label{expmom}
where $\{\l_k\}$ are the eigenvalues of the external field, $\La$. We note
that by means of the relations~\rf{momentderivative1}
and~\rf{momentderivative2}
we can express all the other moments in a similar manner. We note the presence
of cubic singularities which is a well known property of $(3,3m-1)$ and
$(3,3m-2)$ rational matter coupled to gravity. For the $p=n$
version of the Kontsevich model the moment variables depend on $(n-1)$
susceptibilities and
$n$-root singularities are expected.

Let us integrate the equations~\rf{boundary} to obtain the genus zero
contribution to the free energy, $F_0$, which we will need for our
considerations in
the next section.
Exploiting the relations~\rf{momentderivative1} and~\rf{momentderivative2}
and assuming $\frac{dF}{d\th_0}=\frac{dF}{d\tht_0}=0$
for $\th_0=\tht_0=0$ it is easy to show that
\bea
\frac{dF_0}{d\th_0}&=&\frac{3}{2}\left(M_0-u_1 u_2\right), \\
\frac{dF_0}{d\tht_0}&=&3\left(J_0+\frac{1}{3}u_1^3-\frac{1}{2}u_2^2\right)
\eea
and furthermore assuming $F_0=0$ for $\th_0=\tht_0=0$ one arrives at the
following
expression for $F_0$
\beq
F_0=\left(\frac{3}{2}\right)^2\left\{
\frac{1}{2}u_1 u_2^2-u_1J_0-u_2M_0 +M_{-1}+\int J_1 M_1 du_2 \right\}
\label{F0}
\eeq
where in the integral it is understood that $J_1$ and $M_1$ should be expressed
as on the right hand side of~\rf{momentM} and~\rf{momentJ} and $\int$ is short
hand notation for $\int_0^{u_2}$.
This expression can of course be rewritten in the form given in~\cite{EYY94}.
We note that all terms entering~\rf{F0} except the integral $\int J_1 M_1 du_2$
can be expressed in terms of elementary functions of $u_1$, $u_2$ and
$\{\lambda_k\}$.

\newsection{Macroscopic Loops \label{macloop} }
In this section we shall be concerned with the calculation of macroscopic
loops. By macroscopic loops we mean correlation functions of the following type
\beq
W^{(n)}(\pi_1,\ldots,\pi_n)=
\frac{d}{dV(\pi_n)}\ldots\frac{d}{dV(\pi_1)}F
\eeq
where $\frac{d}{dV(\pi)}$, the loop insertion operator, is given by
\beq
\frac{d}{dV(\pi)}=\sum_{k}\left\{\pi^{-k-4/3}\frac{d}{d\th_k}
+\pi^{-k-5/3}\frac{d}{d\tht_k}\right\}.
\eeq
Our aim will be to derive a closed expression for the genus zero contribution
to the $n$-loop correlator, $W_0^{(n)}(\pi_1,\ldots,\pi_n)$.
For that purpose it is
convenient to work with a slightly different version of the loop insertion
operator. Using the boundary equations~\rf{boundary} it is easy to show that
$\frac{d}{dV(\pi)}$ can be rewritten as
\beq
\frac{d}{dV(\pi)}=\frac{\partial}{\partial V(\pi)}+M_2(\pi)\hat{Q}
+J_2(\pi)\hat{P}
\eeq
where
\beq
\frac{\partial}{\partial V(\pi)}
=\sum_k\left\{\pi^{-k-4/3}\frac{\partial}{\partial \th_k}
+\pi^{-k-5/3}\frac{\partial}{\partial \tht_k}\right\}
\eeq
and
\bea
\hat{P}=\Omega_1\frac{\partial}{\partial u_1}+
\Omega_2\frac{\partial}{\partial u_2},
&\hs & \hat{Q}=\Omega_2\frac{\partial}{\partial u_1}-
u_1\Omega_1\frac{\partial}{\partial u_2},
 \\
\Omega_1=\frac{M_2}{(1-J_2)^2+u_1 M_2^2}, &\hs&
\Omega_2=\frac{(1-J_2)}{(1-J_2)^2+u_1 M_2^2},
\eea
\bea
M_k(\pi)= &\frac{\partial M_{k-1}}{\partial V(\pi)}&=
M_k\left|\begin{array}{c}  \\ \lambda_i^3\rightarrow \pi  \end{array}\right.
\label{dMdV}\\
J_k(\pi)=&\frac{\partial J_{k-1}}{\partial V(\pi)}&= J_k\left|\begin{array}{c}
\\ \lambda_i^3\rightarrow\pi  \end{array}\right. \label{dJdV}
\eea
where by $\lambda_i^3\rightarrow \pi$ we mean that the functional dependence of
$M_k(\pi)$ on $\pi$ is like that of $M_k$ on any of the $\lambda_i^3$.
To determine $W_0^{(1)}(\pi)$ we need only to determine
$\frac{\partial F_0}{\partial V(\pi)}$ since as shown in~\cite{EYY94} (and
easily verified for the expression~\rf{F0}) we have
$\frac{\partial F_0}{\partial u_1}=\frac{\partial F_0}{\partial u_2}=0$.
With the notation of equation~\rf{dMdV} and~\rf{dJdV} one finds
\[
W_0^{(1)}(\pi)=
\left(\frac{3}{2}\right)^2\left[
-u_1J_1(\pi)-u_2M_1(\pi)+M_0(\pi)+
\int J_1 M_2(\pi)du_2+\int M_1J_2(\pi) du_2\right].
\]
It is easily verified that
\beq
\frac{\partial}{\partial u_1} W^{(1)}_0(\pi)=\frac{\partial}{\partial u_2}
W^{(1)}_0(\pi)=0.
\eeq
Hence the two-loop correlator reads
\beq
W_0^{(2)}(\pi_1,\pi_2)=\left(\frac{3}{2}\right)^2
\left[\int J_2(\pi_1) M_2(\pi_2) du_2+
\int J_2(\pi_2) M_2(\pi_1)du_2\right].
\label{W02}
\eeq
We note that as expected the two-loop correlator exhibits no explicit
dependence on the time variables. Hence to find the three-loop correlator we
need only to
determine the effect of applying $\frac{\partial}{\partial u_1}$ and
$\frac{\partial}{\partial u_2}$ to $W_0^{(2)}(\pi_1,\pi_2)$. The application of
$\frac{\partial}{\partial u_2}$ is straightforward and by making use of the
relations~\rf{momentderivative1} and~\rf{momentderivative2} one realizes that
the integrals resulting from applying $\frac{\partial}{\partial u_1}$ to the
integrands in~\rf{W02} can actually by carried out by partial integration. The
expression for the three-loop correlator that one arrives at is the following
\bea
W_0^{(3)}(\pi_1,\pi_2,\pi_3)&=&\left(\frac{3}{2}\right)^2\left\{
\Omega_1 J_2(\pi_1) J_2(\pi_2) J_2(\pi_3)-
u_1 \Omega_2 M_2(\pi_1) M_2(\pi_2) M_2(\pi_3)\right. \non \\
&& \mbox{}+\Omega_2\left[M_2(\pi_1)J_2(\pi_2)J_2(\pi_3)+ dis.\; perm.\right]
\non \\
&& \left.\mbox{}-u_1\Omega_1\left[M_2(\pi_1) M_2(\pi_2)
J_2(\pi_3)+dis.perm.\right]
\right\}
\label{W03}
\eea
where here and in the following by $dis.\; perm.$ we mean permutations
of $\pi$'s which give rise to truly different terms.
Before proceeding to the general case let us comment on the geometrical
interpretation of~\rf{W03}
For that purpose let us note that
\bea
\Omega_1=\frac{3}{2} c^3 F_{000},&\hs&\Omega_2=\frac{3}{4}c^2
\tilde{c}\:F_{001}\\
(-u_1)\Omega_1=\frac{3}{8}c\:\tilde{c}^2 F_{011},&\hs&
(-u_1)\Omega_2=\frac{3}{16}\tilde{c}^3 F_{111}
\eea
where
\beq
F_{ijk}=\frac{d^3 F_0}{d t_{i,0}\, dt_{j,0}\, dt_{k,0}}
\eeq
and $c$ and $\tilde{c}$ are given by
\beq
t_{0,0}=c^{-1}\th_0,\hs t_{1,0}=\tilde{c}^{-1}\tilde{\th}_0
\eeq
(cf.\ to equation~\rf{times}).
Hence if we define propagators ${\cal P}^0(\pi)$ and ${\cal P}^1(\pi)$ by
\beq
{\cal P}^0(\pi)=\frac{3}{2} c\: J_2(\pi),\hs
{\cal P}^1(\pi)=\frac{3}{4}\tilde{c}\: M_2(\pi)
\eeq
we can write the three-loop correlator as
\bea
W_0^{(3)}(\pi_1,\pi_2,\pi_3)&= &F_{000}\:\p^0(\pi_1)\p^0(\pi_2)\p^0(\pi_3)+
F_{111}\:\p^1(\pi_1)\p^1(\pi_2)\p^1(\pi_3) \non \\
&&\mbox{}+F_{011}\:\left[\p^0(\pi_1)\p^1(\pi_2)\p^1(\pi_3)+dis.\; perm.\right]
\non \\
&&\mbox{}+F_{001}\:\left[\p^0(\pi_1)\p^0(\pi_2)\p^1(\pi_3)+dis.\; perm.\right]
\label{W03geo}
\eea
and we see that the three-loop correlator is determined by the three-point
vertices of the gravitational primary fields and that $\p^0(\pi)$ and
$\p^1(\pi)$ have a natural interpretation as propagators associated with the
two gravitational primary fields of the model. The formula~\rf{W03geo}
is a natural generalisation of the corresponding formula encountered in the
case of $(p,q)=(2,2m-1)$ minimal models coupled to gravity~\cite{AJM90,MSS91}
and it is natural to expect that the three-loop correlator will have a similar
structure in the generic case. For the series of rational matter fields of the
type $(p,pm-1),\ldots(p,pm-(p-1))$ coupled to gravity  the propagators will
exhibit $p$-root singularities (cf.\ to equations~\rf{dMdV}, \rf{dJdV}, \rf{M0}
and~\rf{J0}.)
The decomposition of the 3-loop correlator into
vertices and propagators is in perfect agreement with the Feynman rules for
calculating multi-loop correlators for unitary conformal models coupled to
2D gravity obtained from the approach of strings with discrete target
space~\cite{Kos91}.

Let us proceed now to the general case. To calculate the $n$-loop correlator
$(n>3)$ we must apply the loop insertion operator $(n-3)$ times to each of the
terms in equation~\rf{W03}. The result of this process can be given in a closed
form. For instance
\bea
\lefteqn{
\frac{d}{dV(\pi_{n+3})}\frac{d}{dV(\pi_{n+2})}\ldots\frac{d}{dV(\pi_4)}
\left\{\Omega_1 J_2(\pi_1) J_2(\pi_2) J_2(\pi_3)\right\}=}\non \\
&&\left\{\sum_{k=0}^n \hat{P}^k \hat{Q}^{n-k} \Omega_1
\left[\overbrace{J_2(\pi_4)\ldots J_2(\pi_{k+4})}^{k\;terms}
\overbrace{M_2(\pi_{k+5})\ldots M_2(\pi_{n+3})}^{(n-k)\;terms}+dis.\;
perm.\right] \right.\non\\
&&+ \sum_{k=0}^{n-1}\hat{P}^k\hat{Q}^{n-k-1}
\left(u_1 \Omega_1^2+\Omega_2^2\right)\times \non\\
&&\left.\left[
\frac{\partial M_2(\pi_{n+3})}{\partial u_2}
\overbrace{J_2(\pi_4)\ldots J_2(\pi_{k+4})}^{k\;terms}
\overbrace{M_2(\pi_{k+5})\ldots M_2(\pi_{n+2})}^{(n-k-1)\;terms}
+dis.\;perm. \right]\right\}\times \non\\
&&J_2(\pi_1)J_2(\pi_2)J_2(\pi_3). \label{nloop}
\eea
Here $J_2(\pi_1)J_2(\pi_2)J_2(\pi_3)$ can be replaced by any function
$f(u_1,u_2)$ with no explicit dependence on the time variables
$\{\th_i,\tht_i\}$. In particular the result immediately generalizes to the
case where the loop insertion operator acts on the last term in~\rf{W03}.
If one has in the first line of~\rf{nloop} in stead of a function of the type
$\Omega_1f(u_1,u_2)$ a function of the type $\Omega_2 f(u_1,u_2)$ the formula
still holds provided on the right hand side in the first line $\Omega_1$ is
replaced by $\Omega_2$ and in the second line
$\frac{\partial M_2(\pi)}{\partial u_2}$ is replaced by $-\frac{\partial
J_2(\pi)}{\partial u_2}$. Collecting these facts one can easily write down a
closed expression for the full $(n+3)$-loop correlator. We shall refrain from
doing so. That the stated form of the $(n+3)$-loop correlator
is indeed correct can be proven by induction (generalizing the idea of
reference~\cite{AJM90}) using the following identities
\beq
\left[\hat{P},\hat{Q}\right]=0
\eeq
and
\bea
\left[\frac{\partial}{\partial V(\pi)},\hat{P}^n \hat{Q}^m \Omega_1\right]&=&
\hat{P}^n\hat{Q}^{m+1}\Omega_1 M_2(\pi) +
\hat{P}^{n+1}\hat{Q}^m\Omega_1 J_2(\pi) \non \\
&&\mbox{}-M_2(\pi)\hat{P}^n\hat{Q}^{m+1}\Omega_1-
J_2(\pi)\hat{P}^{n+1}\hat{Q}^m\Omega_1 \non \\
&&
+\hat{P}^n\hat{Q}^m\left(u_1\Omega_1^2+\Omega_2^2\right)
\frac{\partial M_2(\pi)}{\partial u_2},
\eea
\bea
\left[\frac{\partial}{\partial V(\pi)},\hat{P}^n \hat{Q}^m \Omega_2\right]&=&
\hat{P}^n\hat{Q}^{m+1}\Omega_2 M_2(\pi) +
\hat{P}^{n+1}\hat{Q}^m\Omega_2 J_2(\pi) \non \\
&&\mbox{}-M_2(\pi)\hat{P}^n\hat{Q}^{m+1}\Omega_2-
J_2(\pi)\hat{P}^{n+1}\hat{Q}^m\Omega_2 \non \\
&&
-\hat{P}^n\hat{Q}^m\left(\Omega_2^2+u_1\Omega_1^2\right)
\frac{\partial J_2(\pi)}{\partial u_2},
\eea
\bea
\lefteqn{\left[\frac{\partial}{\partial V(\pi)},\hat{P}^n \hat{Q}^m
\left(u_1\Omega_1^2+\Omega_2^2\right)\right]=}\non\\
&&
\hat{P}^n\hat{Q}^{m+1}\left(u_1\Omega_1^2+\Omega_2^2\right)  M_2(\pi) +
\hat{P}^{n+1}\hat{Q}^m \left(u_1\Omega_1^2+\Omega_2^2\right) J_2(\pi) \non \\
&&\mbox{}-M_2(\pi)\hat{P}^n\hat{Q}^{m+1}\left(u_1\Omega_1^2+\Omega_2^2\right)
-J_2(\pi)\hat{P}^{n+1}\hat{Q}^m \left(u_1\Omega_1^2+\Omega_2^2\right)
\eea
The three last relations themselves can likewise be proven by induction.  The
only non-standard part is to realize that the following equations hold
\bea
\frac{\partial \Omega_1}{\partial V(\pi)} &=&
\Omega_1\left(\hat{P}J_2(\pi)\right)+
\Omega_2\left(\hat{P}M_2(\pi)\right), \\
\frac{\partial \Omega_2}{\partial V(\pi)} &=&
(-u_1)\Omega_1\left(\hat{P}M_2(\pi)\right)
+\Omega_2\left(\hat{P}J_2(\pi)\right)
 \non\\
&=&
\Omega_2\left(\hat{Q}M_2(\pi)\right)
+\Omega_1\left(\hat{Q}J_2(\pi)\right).
\eea

It is easy to see that among the contributions to the
$n$-loop correlator ($n>3$)
we have terms of the same type as those constituting the 3-loop correlator,
namely terms constisting of the $n$-point vertices of the gravitational primary
fields saturated by propagators, ${\cal P}^0(\pi)$ and ${\cal P}^1(\pi)$. This
follows from the following observation
\bea
\hat{P}f=\left(\frac{3}{2}c\right)\frac{df}{dt_{0,0}}
&\Longleftrightarrow &\frac{\partial f}{\partial t_{0,0}}=0,\non \\
\hat{Q}f=\left(\frac{3}{4}\tilde{c}\right)\frac{df}{dt_{1,0}}&
\Longleftrightarrow& \frac{\partial f}{\partial t_{1,0}}=0 \non
\eea
and the fact that neither $\Omega_1$ nor $\Omega_2$ has any explicit dependence
on $\th_0$ or $\tht_0$ (cf.\ to~\rf{momentM} and~\rf{momentJ}).
We would of course expect terms of the type just mentioned to be present for
any
series of rational matter fields coupled to gravity.
The terms which
are not of this type all contain products of (at most $(n+1)$) $m$-point
vertices with $3\leq m\leq n-1$. It would be interesting to disentangle the
$\pi$-dependent factors in these terms to obtain an interpretation of these in
terms of internal and external propagators in the spirit of the theory of
strings with discrete target space~\cite{Kos91}. This would provide us with an
expression which could be immediately generalized to the case of the generic
rational matter field.

Let us close this section be remarking that the formula~\rf{nloop} is a little
more involved than one could have hoped for knowing the corresponding formula
for the case of minimal models with $(p,q)=(2,2m-1)$ coupled to
gravity~\cite{AJM90,MSS91}. In the latter case the expression for the
$(n+3)$-loop correlator consists of only one term with a structure similar to
that of the first term on the right hand side of~\rf{nloop}. However, for a
model with the presence of two operators with different dimensions we must
accept a less simple result.

\newsection{Outlook \label{conclusion} }
Using the moment description of the generic 1-matrix model~\cite{ACKM93} it was
proven that the free energy of the Kontsevich model was exactly equal to that
of
the generic 1-matrix model with all subleading singularities
subtracted~\cite{AK93}. It would be interesting to establish a similar
correspondence between the two matrix model and the generalized Kontsevich
models. The $p=3$ version, that we have considered here, we would expect to
have a singularity structure describing the leading singularities of a
two-matrix model where one  matrix potential is cubic and the other one
arbitrary. This is
of course in accordance with the fact that the coupling to gravity of all
rational matter fields of the type $(3,3m-1)$, $(3,3m-2)$ can be described by
a two-matrix model of the type mentioned~\cite{DKK93}. The correspondence is
furthermore outlined by the following fact. For a two-matrix model with one
potential cubic the loop equation giving the 1-loop correlator of the  matrix
with the arbitrary potential (an algebraic equation of degree 3~\cite{cubic})
has exactly the same structure as the matrix Airy equation satisfied by
$Z_3^N(\La)$~\cite{IZ92}. Likewise we would expect the $p=n$ version of the
Kontsevich model to give exactly the leading singular behaviour of a two matrix
model with one potential of degree $n$ and the other
one arbitrary. A moment description of the two-matrix model has not yet been
found but should certainly exist. In the light of the discussion above a
reasonable strategy for finding such a description would be to start by
considering a two-matrix model with one potential cubic. One would expect more
than two series of moments to be necessary in analogy with the 1-matrix  case
where one set of moments was sufficient for the double scaling limit but two
sets were needed away from this limit. Likewise on the basis of the experience
from the 1-matrix model one might expect complications to occur at genus zero.
Finding the appropriate moment description for the two matrix model would,
however, provide us with the exact correspondence between the matrix model
coupling constants and the continuum time variables used in the context of
topological gravity which again would allow us to understand the connection
between the matrix model observables and the continuum scaling operators.

Furthermore it would be interesting to elaborate on the correspondence between
the present approach and the approach of strings with discrete target
space~\cite{Kos91} in order to obtain a geometrically more appealing version
of~\rf{nloop} as well as a generalization thereof to arbitrary $(p,q)$ rational
matter fields coupled to gravity.

\vspace{12pt}
\noindent
{\bf Acknowledgements}\hspace{0.3cm}
It is a pleasure to thank P.\ Di Francesco, C.\ Itzykson and I.\ Kostov for
interesting and stimulating discussions.

\end{document}